\documentclass[12pt,superscriptaddress,showpacs,showkeys,onecolumn,prb,aps,a4paper,floatfix]{revtex4-1}
\usepackage{rotating}
\usepackage{amssymb}
\usepackage{mathptmx}

\usepackage{esvect}
\usepackage{graphicx}
\graphicspath{{figures/}}
\usepackage{floatrow}

\usepackage{natbib}
\usepackage{physics}
\usepackage[para,online,flushleft]{threeparttable} 
\usepackage{array}
\usepackage{ragged2e} 

\makeatletter
\newcommand*{\citenst}[2][]{%
  \begingroup
  \let\NAT@mbox=\mbox
  \let\@cite\NAT@citenum
  \let\NAT@space\NAT@spacechar
  \let\NAT@super@kern\relax
  \renewcommand\NAT@open{[}%
  \renewcommand\NAT@close{]}%
  \citet[#1]{#2}%
  \endgroup
}
\makeatother

\makeatletter
\newcommand*{\citenumns}[2][]{%
  \begingroup
  \let\NAT@mbox=\mbox
  \let\@cite\NAT@citenum
  \let\NAT@space\NAT@spacechar
  \let\NAT@super@kern\relax
  \renewcommand\NAT@open{[}
  \renewcommand\NAT@close{]}%
  \cite[#1]{#2}
  \endgroup
}
\makeatother
\usepackage[caption=false]{subfig}
\makeatletter

\bibpunct{}{}{,}{s}{}{,}

\begin{document}

\newcommand{\hdblarrow}{H\makebox[0.9ex][l]{$\downdownarrows$}-}
\title{Title}

\title{Nonlinear Properties of Supercurrent-Carrying Single and Multi-Layer Thin-Film Superconductors}
\author{Songyuan Zhao}
\email{sz311@cam.ac.uk}
\author{S. Withington}
\author{D. J. Goldie}
\author{C. N. Thomas}
\affiliation{Cavendish Laboratory, JJ Thomson Avenue, Cambridge CB3 OHE, United Kingdom.}
\date{12-December-2019}

\begin{abstract}
  Superconducting thin-films are central to the operation of many kinds of quantum sensors and quantum computing devices: Kinetic Inductance Detectors (KIDs), Travelling-Wave Parametric Amplifiers (TWPAs), Qubits, and Spin-based Quantum Memory elements. In all cases, the nonlinearity resulting from the supercurrent is a critical aspect of behaviour, either because it is central to the operation of the device (TWPA), or because it results in non-ideal second-order effects (KID).

  Here we present an analysis of supercurrent carrying superconducting thin-films that is based on the generalized Usadel equations. Our analysis framework is suitable for both homogeneous and multilayer thin-films, and can be used to calculate the resulting density of states, superconducting transition temperature, superconducting critical current, complex conductivities, complex surface impedances, transmission line propagation constants, and nonlinear kinetic inductances in the presence of supercurrent. Our analysis gives the scale of kinetic inductance nonlinearity (I*) for a given material combination and geometry, and is important in optimizing the design of detectors and amplifiers in terms of materials, geometries, and dimensions.

  To investigate the validity of our analysis across a wide range of supercurrent, we have measured the transition temperatures of superconducting thin-films as a function of DC supercurrent. These measurements show good agreement with our theoretical predictions in the experimentally relevant range of current values.
\end{abstract}

\keywords{kinetic inductance, nonlinearity, Usadel equations, parametric amplifiers}
\maketitle

\section{Introduction}
Owing to their low-loss, high quality factor characteristics below their superconducting transition temperatures ($T_c$), superconducting thin-films are important to the operation of many kinds of quantum sensors and quantum computing devices, such as Kinetic Inductance Detectors (KIDs) \citenumns{Day_2003}, Travelling-Wave Parametric Amplifiers (TWPAs) \citenumns{Eom_2012}, Kinetic Inductance Parametric Up-Converters (KPUPs) \citenumns{Kher_2016}, Superconducting Qubits \citenumns{Ferguson_2006}, and Spin-based Quantum Memory elements \citenumns{Parkin_2003, Bienfait_2015}. When designing these superconducting devices, an important consideration is the nonlinearity in superconducting kinetic inductance with respect to supercurrent \citenumns{Pippard_1950,Pippard_1953}. The nonlinear inductance of a superconducting device is expected to have the form \citenumns{Jonas_review}
\begin{align}
  L=L_0\left(1+\frac{I^2}{I_*^2}+...\right) \,,  \label{eq:nonlinear_basic}
\end{align}
where $L$ is the inductance of the device, $L_0$ is the inductance in the absence of supercurrent, $I$ is the supercurrent, and $I_*$ is the scale of the quadratic inductance nonlinearity.

In the case of TWPAs and KPUPs, this nonlinear kinetic inductance is critical to the operation and performance of the devices \citenumns{Eom_2012,Voronin_1979,Chaudhuri_2015,Songyuan_2019_paramp,Kher_2016}; in other cases, the nonlinear kinetic inductance results in non-ideal behaviour that is important even in common device operation power regimes, which often involve high readout power in order to improve noise performance \citenumns{Jonas_review}. As such, understanding and calculation of the nonlinear kinetic inductance is important to the quantitative design processes of these thin-film devices.

In the past decade, there has also been considerable research in applying DC bias current across high quality superconducting thin-films \citenumns{Chen_2011,Li_2013,Hao_2014,Bosman_2015,Vissers_2015,Adamyan_2016}, in order to facilitate circuit quantum electrodynamics experiments \citenumns{Chen_2011,Li_2013,Hao_2014,Bosman_2015} and to improve versatility associated with frequency tuneability \citenumns{Vissers_2015,Adamyan_2016}.

Analyses of supercurrent in superconducting thin-films can be based on the Usadel equations, which is a set of diffusive-limit equations derived from the Bardeen–Cooper–Schrieffer (BCS) theory of superconductivity \citenumns{Usadel1970,Anthore_2003,Clem_2012}. Anthore et al. have calculated and experimentally measured the resultant density of states in a superconducting thin-film due to supercurrent using the Usadel equations \citenumns{Anthore_2003}. The theory and experiment demonstrated excellent agreement, lending confidence to the use of the Usadel equations as the foundation of our analysis framework. Further, the work by Clem et al. \citenumns{Clem_2012} based on the Usadel equations has been applied experimentally to estimate the depairing current of superconducting nanowires to good agreement \citenumns{Frasca_2019}.  The paper by Anthore et al. in particular presents a series expansion of the superconducting order parameter ($\Delta$) with respect to supercurrent for single layer superconducting thin-films. This series expansion has been used by other studies to estimate the superconductor complex conductivities and kinetic inductances \citenumns{Jonas_review, Kher_2017}. As we shall demonstrate in this study, this approximate approach does not account for the change in the shape of the density of states, and underestimates the impact of supercurrent.

Using the full density of states as an input to Nam's equations \citenumns{Nam_1967}, we compute the complex conductivities of the thin-films. We then compute the surface impedances using the transfer matrix method \citenumns{Songyuan_2018}. Finally, we calculate the transmission line inductances from the surface impedances by using the appropriate transmission line theory for the geometry of the device \citenumns{Songyuan_transmission_lines_2018}, such microstrip transmission line or coplanar waveguide.

We have also measured the supercurrent dependence of the superconducting transition temperatures for single-layer titanium (Ti) and multi-layer aluminium-titanium (Al-Ti) thin films. Our results confirm the validity of the Usadel theory approach for experimentally realistic device dimensions and current regimes.

\section{Theory}

\subsection{Usadel equations}
In this analysis, the multilayers are stacked in the $x$ direction, and the supercurrent flows in the $z$ direction.
\label{sec:Usadel}
The Usadel equations in one dimension are \citenumns{Anthore_2003,Usadel1970,Brammertz2004,Golubov2004,Vasenko2008}
\begin{equation}
\label{eq:usadel_a}
\frac{\hbar D_S}{2} \nabla^2{\theta}+iE \sin \theta + \Delta \cos \theta - \frac{\hbar}{2D_S}\vv{v}_s^{\,2}\cos\theta\sin\theta = 0 ,
\end{equation}
and
\begin{equation}
\label{eq:selfCon_a}
\Delta=N_SV_{0,S}\int^{k_B\Theta_{D,S}}_0 dE \operatorname{tanh}\left(\frac{E}{2k_BT}\right)\operatorname{Im}\left(\sin\theta\right) ,
\end{equation}
where $\theta$ is a complex variable dependent on energy $E$ parametrising the superconducting properties, $N_S$ is the electron single spin density of states, $V_{0,S}$ is the superconductor interaction potential, $\Delta$ is the superconductor order parameter, $k_B\Theta_{D,S}$ is the Debye energy, $k_B$ is the Boltzmann constant, $T$ is the temperature of the superconducting film, $D_S$ is the diffusivity constant, given by $D_S=\sigma_N/(N_Se^2)$ \citenumns{Martinis_2000}, $e$ is the elementary charge, $\operatorname{Im}(x)$ takes the imaginary part of $x$, and finally $\sigma_N$ is the normal state conductivity, at $T$ just above $T_c$. Equation~(\ref{eq:selfCon_a}) is the self-consistency equation for order parameter $\Delta$. We have introduced the superfluid velocity $\vv{v}_s=D_S[\vv{\nabla}\phi-(2e/\hbar)\vv{A}]$, where $\phi$ is the superconducting phase, and $\vv{A}$ is the magnetic vector potential. We assume that the effect due to the induced field is negligible compared to that of supercurrent. \citenumns{Anthore_2003}

The supercurrent density $\vv{j}$ is given by
\begin{align}
  \vv{j} &= \frac{\sigma_N}{eD_S}\int_0^\infty dE\,\tanh\left(\frac{E}{2k_BT}\right)\operatorname{Im}(\sin^2\theta)\vv{v}_s\,.
\end{align}
For supercurrent flowing in the $z$-direction, $\vv{v}_s^{\,2}=D_S^2\left(\partial\phi/\partial z\right)^2$. For the case of a homogeneous BCS superconductor, the first term of equation~(\ref{eq:usadel_a}) can be removed, simplifying equation~(\ref{eq:usadel_a}) into
\begin{equation}
\label{eq:usadel_homo}
iE \sin \theta + \Delta(x) \cos \theta - \Gamma\cos\theta\sin\theta = 0 ,
\end{equation}
where $\Gamma = \hbar D_S/2*(\partial\phi/\partial z)^2$ is the depairing factor. The above equation can be solved iteratively with equation~(\ref{eq:selfCon_a}) to obtain $\Delta(\Gamma)$. Numerically, it is easier to solve equation~(\ref{eq:usadel_homo}) for $\sin(\theta)$ using a polynomial root finder, rather than finding $\theta$ directly.


In the case of a multilayer superconductor, the boundary conditions (BCs) between the layers need to be taken into account. The BCs suitable for the Usadel equations can be found in \citenumns{Songyuan_2018}. Instead of calculating nonlinearity with respect to $\Gamma$, which is not constant across the multilayer, calculations should be performed with respect to $\partial \phi/\partial z$. $\partial \phi/\partial z$ cannot vary across the multilayer (in the $x$ direction) due to the absence of net supercurrent (in the $x$ direction). Computation-time-wise, it is beneficial to adopt the thin-film approximation scheme that has demonstrated good agreement with experiment for multilayer superconductors. The approximation assumes $\theta$ varies slowly, and can be accounted by a second order polynomial expansion. \citenumns{Martinis_2000, Songyuan_2018_Tc}

\subsection{Complex Conductivities and Impedances}
Nam's equations \citenumns{Nam1967} are a generalization of the Mattis-Bardeen \citenumns{MattisBardeen} theory into strong-coupling and impure superconductors. Nam's equations compute the complex conductivity $\sigma = \sigma_1-j\sigma_2$ using a pair of integrals of $\theta$ across energy $E$. The integrals, as well as their evaluations for Al-Ti bilayers can be can be found in \citenumns{Songyuan_2018}. 
After calculating $\sigma$, the complex surface impedance for a homogeneous single layer can then be obtained using \citenumns{Kautz_1978}
\begin{align}
Z_s=\left(\frac{j\omega\mu_0}{\sigma}\right)^{1/2}\operatorname{coth}[(j\omega\mu_0\sigma)^{1/2}t] ,
\end{align}
where $t$ is the thickness of the homogeneous superconducting film, and $\mu_0$ is the vacuum permeability.

For multilayers, $Z_s$ can be found by dividing the multilayer into thin layers of thickness $\delta x$, and then cascading the resultant transfer matrices along the multilayer. A detailed discussion of the above methodology, as well as an analysis of numerical results for Al-Ti multilayers, can be found in \citenumns{Songyuan_2018}.

\subsection{Transmission Line Properties}
The series impedance and shunt admittance of a transmission line structure can be calculated from $Z_s$ as follows \citenumns{Whitaker_1988,Withington_1995,Songyuan_transmission_lines_2018}:
\begin{align}
\label{eqn:Z}
Z &=j(k_0\eta_0)g_1 +2\sum_{n}g_{2,n}Z_{s,n} \\
\label{eqn:Y}
Y &=j\left(\frac{k_0}{\eta_0}\right)\left(\frac{\epsilon_{fm}}{g_1}\right) \, ,
\end{align}
where $k_0$ is the free-space wavenumber, $\eta_0$ is the impedance of free-space, subscript $n$ identifies superconductor surfaces, which are upper, lower, and ground surfaces, denoted by subscripts $u,l,$ and $g$ respectively, $\epsilon_{fm}$ is the effective modal dielectric constant, which is given by existing normal conductor transmission line theories, for example \citenumns{Edwards_1976, Gupta_1996}. $g_1$ and $g_2$ are geometric factors which can be calculated using appropriate conformal mapping theories \citenumns{Withington_1995,Songyuan_transmission_lines_2018}.

After obtaining the series impedance and the shunt admittance, other properties of the superconducting transmission line can be calculated straightforwardly. The characteristic impedance is given by $\eta = (Z/Y)^{1/2}$. The propagation constant is given by $\gamma=\alpha+j\beta=({ZY})^{1/2}$, where $\alpha$ is the attenuation constant and $\beta$ is the phase constant. The inductance per unit length $L$ can finally be calculated using $L=\operatorname{Im}(Z)/\omega$. The calculation can then be iterated for different values of $I$ to obtain $L(I)$, which allows the extraction of $I_*$ using a polynomial fit.

\section{Results and Discussions}


\begin{figure}[ht]
\includegraphics[width=1.0\textwidth]{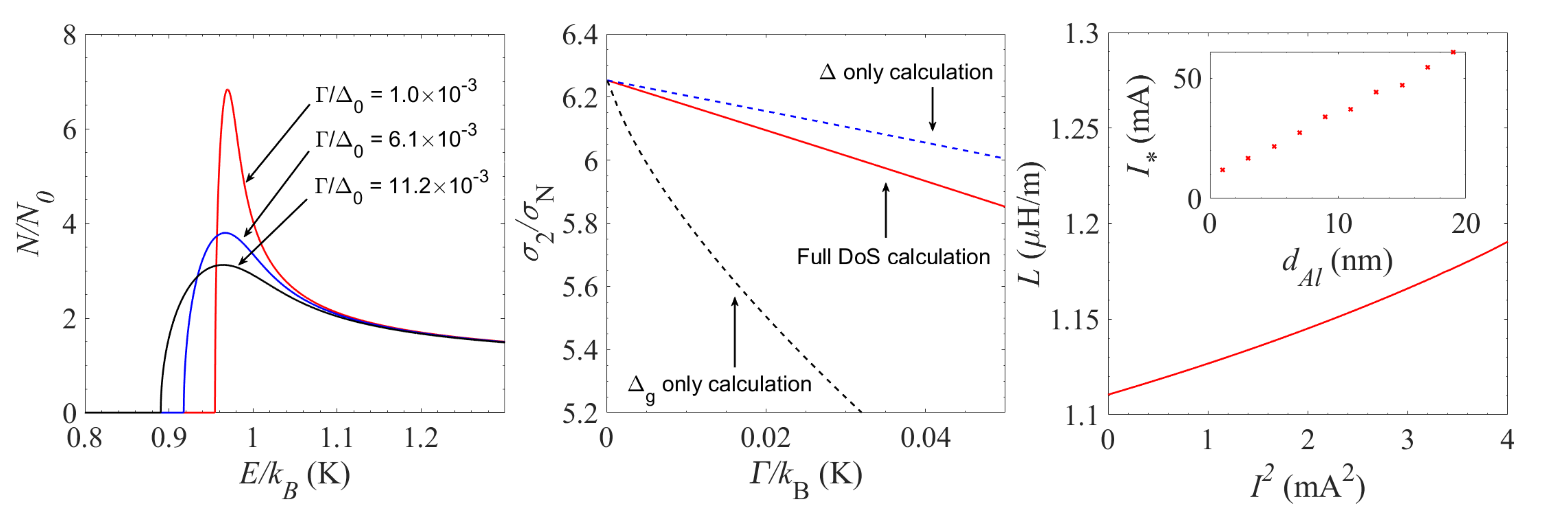}
\caption{\label{fig:Combined_Fig_1} Left figure: Plot of Ti superconducting density of states $N/N_0$ against energy $E/k_{B}$ at temperature $T=0.01\,\mathrm{K}$ for different values of supercurrent depairing factor $\Gamma/\Delta_0$. Red line: $\Gamma/\Delta_0 = 1.0\times10^{-3}$, $1-\Delta/\Delta_0 = 0.6\times10^{-3}$; blue line: $\Gamma/\Delta_0 = 6.1\times10^{-3}$, $1-\Delta/\Delta_0 = 4.5\times10^{-3}$; black line: $\Gamma/\Delta_0 = 11.2\times10^{-3}$, $1-\Delta/\Delta_0 = 8.4\times10^{-3}$. Middle figure: Plot of normalized reactive conductivity $\sigma_2/\sigma_N$ against supercurrent depairing factor $\Gamma/k_{B}$ for Ti at temperature $T=0.01\,\mathrm{K}$, frequency $f=10\,\mathrm{GHz}$. Red line: calculation performed by solving Nam's equations using the full densities of states; blue line: calculation performed using a simplified density of states replacing $\Delta_0$ with suppressed superconducting order parameter $\Delta(\Gamma)$; black line: calculation performed using a simplified density of states replacing $\Delta_0$ with suppressed superconducting DoS gap $\Delta_g(\Gamma)$. Right figure: Plot of inductance per unit length $L$ against squared supercurrent $I^2$ for a Ti microstrip line with thickness $t=100\,\mathrm{nm}$, width $w=5\,\mathrm{\mu m}$, dielectric height $h=250\,\mathrm{nm}$, ground plane Ti thickness $t_g=200\,\mathrm{nm}$, at temperature $T=0.01\,\mathrm{K}$, frequency $f=10\,\mathrm{GHz}$. Inset: Plot of inductance nonlinearity factor $I_*$ against Al thickness $t_{Al}$ for a bilayer Al-Ti microstrip with Ti thickness $t_{Ti}=100\,\mathrm{nm}$, width $w=5\,\mathrm{\mu m}$, dielectric height $h=250\,\mathrm{nm}$, ground plane Ti thickness $t_g=200\,\mathrm{nm}$, at temperature $T=0.01\,\mathrm{K}$. }
\end{figure}


The left figure of Fig.~\ref{fig:Combined_Fig_1} shows Ti superconducting density of states (DoS) $N/N_0=\operatorname{Re}[\cos{\theta}]$ against energy $E/k_{B}$ at temperature $T=0.01\,\mathrm{K}$ for different values of $\Gamma/\Delta_0$, where $\Delta_0\approx1.764 \,k_B T_c$ is the superconducting energy gap of Ti in the absence of supercurrent. The presence of supercurrent broadens the DoS. This is a real effect and it has been experimentally observed by \citenumns{Anthore_2003}. Previous approximations on the inductance nonlinearity \citenumns{Jonas_review, Kher_2017} assume the effect of this new DoS on conductivity can be approximated by a single parameter $\Delta$. Effectively, these studies have assumed that the new DoS can be approximated by a zero-supercurrent DoS shifted to an altered DoS gap at $\Delta$. For convenience we label this simplified DoS function as $n_{j=0}[\Delta(\Gamma)]$. As we see in the middle figure, this assumption leads to underestimation on the impact of the supercurrent.


The middle figure of Fig.~\ref{fig:Combined_Fig_1} shows a plot of normalized reactive conductivity $\sigma_2/\sigma_N$ against $\Gamma/k_{B}$ for Ti at $T=0.01\,\mathrm{K}$, frequency $f=10\,\mathrm{GHz}$. The red line shows calculation performed by solving Nam's equations using the full densities of states shown in left figure. The blue line shows calculation performed using $n_{j=0}[\Delta(\Gamma)]$. The black line shows calculation performed using $n_{j=0}[\Delta_g(\Gamma)]$, where $\Delta_g$ is the energy at which the broadened DoS becomes non-zero. Since $\hbar\omega\ll\Delta_0$, the blue and black lines have approximate forms $\sigma_2/\sigma_N=(\pi\Delta)/(\hbar\omega)$ and $\sigma_2/\sigma_N=(\pi\Delta_g)/(\hbar\omega)$ respectively. Comparing the red line with the blue line, we notice that approximation using $n_{j=0}[\Delta(\Gamma)]$ underestimates the effect of supercurrent. This shows that the broadened DoS in the presence of supercurrent cannot be approximated well using a single energy parameter $\Delta(\Gamma)$. Comparing the red line with the black line, approximation using $n_{j=0}[\Delta_g(\Gamma)]$ overestimates the effect of supercurrent. This is because, in the presence of supercurrent, the DoS is broadened. As a result, $\Delta_g$ shifts further than the overall DoS. The above results highlight the need to perform the full calculation as detailed in this manuscript. For practical purposes, we give the approximation of $\sigma_2/\sigma_N$ as a function of $\Gamma/\Delta_0$, valid for $k_B T \ll \Delta_0$, $\hbar\omega\ll2\Delta_0$, and $\Gamma/\Delta_0<0.2$:
\begin{align}
\frac{\sigma_2}{\sigma_N} =\frac{\pi\Delta_0}{\hbar\omega}\left[1-1.2 \left(\frac{\Gamma}{\Delta_0}\right) -0.50\left(\frac{\Gamma}{\Delta_0}\right)^2 \right]\,.
\end{align}
This expansion can be used in conjunction with previous results from \citenumns{Anthore_2003}, which states that the supercurrent is given by $I=I_{\Gamma}\sqrt{\Gamma/\Delta_0}U_S/\Delta_0$, where $I_{\Gamma}=\sqrt{2}S\Delta_0\sigma_N/(e\xi)$, $S$ is the cross-sectional area, $\xi$ is the superconducting coherence length, $e$ is the electron charge, and $U_S$ can be approximated by:
\begin{align}
\frac{U_S}{\Delta_0} =\frac{\pi}{2}-1.8\left(\frac{\Gamma}{\Delta_0}\right)-1.0\left(\frac{\Gamma}{\Delta_0}\right)^2\,.
\end{align}


The right figure of Fig.~\ref{fig:Combined_Fig_1} shows a plot of inductance per unit length $L$ against squared supercurrent $I^2$ for a Ti microstrip line with thickness $t=100\,\mathrm{nm}$, width $w=5\,\mathrm{\mu m}$, dielectric height $h=250\,\mathrm{nm}$, ground plane Ti thickness $t_g=200\,\mathrm{nm}$. We see from the figure that $L$ can be approximated well by a quadratic expansion on $I$ at small current values. At larger values, an additional quartic term is needed to encapsulate the superconductor response:
\begin{align}
  L=L_0\left(1+\frac{I^2}{I_*^2}+\frac{I^4}{I_{*,4}^4}\right) \,,  \label{eq:nonlinear_quartic}
\end{align}
where $I_{*,4}$ is the scale of the quartic order of inductance nonlinearity. The Ti microstrip studied here has $I_*=8.5\,\mathrm{mA}$ and $I_{*,4}=5.5\,\mathrm{mA}$.
 Inset of the right figure of Fig.~\ref{fig:Combined_Fig_1} shows a plot of $I_*$ against Al thickness $t_{Al}$ for a bilayer Al-Ti microstrip with fixed Ti thickness $t_{Ti}=100\,\mathrm{nm}$, width $w=5\,\mathrm{\mu m}$, dielectric height $h=250\,\mathrm{nm}$, ground plane Ti thickness $t_g=200\,\mathrm{nm}$. As $t_{Al}$ increases, the nonlinear behaviour of the microstrip decreases in significance: this is reflected in the higher $I_*$ values. This trend agrees with our expectations: the presence of an Al layer decreases the resistivity of the multilayer. This lower resistivity in turn results in smaller nonlinearity \citenumns{Eom_2012,Kher_2017}.


\section{Critical Temperature Experiment}
Many aspects of our analysis routine have been individually experimentally established by previous studies: the analysis of superconducting multilayers using the Usadel equations has been justified by \citenumns{Brammertz_2002, Songyuan_2018_Tc}; the analysis of supercurrent using the Usadel equations has been justified by \citenumns{Anthore_2003}; the computation of complex conductivities using Nam's equations has been justified by \citenumns{Nam_1967}; the calculation of transmission line properties using conformal mapping analysis has been justified by \citenumns{Visser_2014,Yassin_2000,Shan_2007}.

Despite the above experimental justifications, a caveat exists regarding the analysis of real superconducting devices using the Usadel equations: the physical dimensions of the devices tested in previous studies have physical dimensions smaller than the relevant length scales of the material system, i.e. the perpendicular field penetration depth $\lambda_{\perp}$ and the superconducting coherence length $\xi$ as identified by \citenumns{Romijn_1982}. To illustrate, the aluminium strip tested by \citenumns{Anthore_2003} has width $w=120\,\mathrm{nm}$ and thickness $t=40\,\mathrm{nm}$; the aluminium strips tested by \citenumns{Romijn_1982} has dimensions $w=30-61\,\mathrm{nm}$, $t=20-89\,\mathrm{nm}$. These dimensions are much smaller than that typically used in the design of KIDs and TWPAs, and many real devices have dimensions comparable to, or exceeding, one or both of the length scales. For a thin-film with $w>\lambda_{\perp}$, the supercurrent distribution becomes non-uniform as current piles up near the edges \citenumns{Tinkham_1994}; for a thin-film with $w>4.4\xi$, at high current densities, vortex formation will result in deviations from ideal behaviour \citenumns{Likharev_1979}. The thin-film parallel field penetration depth is given by $\lambda_{\parallel} \approx \lambda_L\sqrt{{\xi_0}/{l}}$, the thin-film perpendicular field penetration depth is given by $\lambda_{\perp}=\lambda_{\parallel}^2/t$, and the low temperature coherence length is given by $\xi\approx\sqrt{\xi_0 l}$ where $\xi_0=\hbar v_F/(\pi\Delta_0)$ is the bulk coherence length, $l$ is the mean free path, $v_F$ is the Fermi velocity, $\lambda_L=\sqrt{m_e/(\mu_0 n e^2)}$ is the London depth, $m_e$ is the electron mass, and $n$ is the electron density. \citenumns{Romijn_1982,Tinkham_1994,Anthore_2003} For our $25\,\mathrm{nm}$ aluminium thin-films, $\lambda_{\perp,Al}=0.40\,\mathrm{\mu m}$, $\xi_{Al}=0.20\,\mathrm{\mu m}$; for our $100\,\mathrm{nm}$ titanium thin-films, $\lambda_{\perp,Ti}=0.12\,\mathrm{\mu m}$, $\xi_{Ti}=0.56\,\mathrm{\mu m}$. Here we have used data for Al and Ti properties from \citenumns{Songyuan_2018_Tc}, supplemented by mean free path data from \citenumns{Day_1995,Gall_2016}. For superconducting strips with $w$ on the order of a few microns \citenumns{Doyle_2008,Eom_2012,Jiansong_2012,Li_2013,Vissers_2015,Shan_2016} on the border of the relevant length scales, it is useful to determine the range of current within which the 1D Usadel equations treatment of the supercurrent provides a good prediction of device behaviour. To this end, we have performed an experiment measuring the $T_c$ of a superconducting strip for a given supercurrent $I$.

Ti and Al-Ti films were deposited by DC magnetron sputtering at a base pressure of $2 \times 10^{-10}$ Torr or below. For bilayer films, Al layers were deposited after Ti layers without breaking the vacuum. The films were patterned to achieve four-terminal sensing geometry and connected to electronics via Al wire-bonds. The samples were mounted to the cold stage of a dilution refrigerator inside a niobium magnetic shield. Temperature monitoring was achieved using a calibrated ruthenium oxide thermometer. For each set of measurement, a fixed current was first injected to the mounted superconducting film. The temperature of the cold stage was then slowly raised until transition from superconducting to normal state had occurred. The potential difference across the film was continuously measured throughout this transition process.

\begin{figure}[ht]
\includegraphics[width=1.0\textwidth]{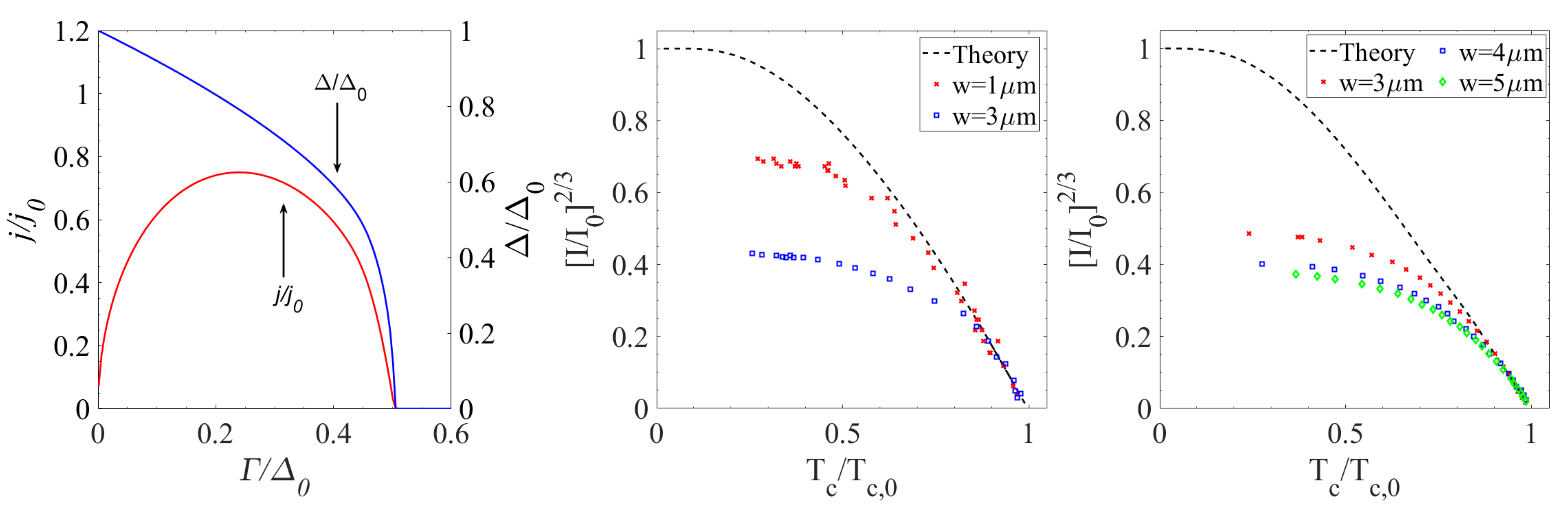}
\caption{\label{fig:Combined_Fig_2} Left figure: Red line, left axis: plot of Ti scaled current density $j/j_0$ against scaled supercurrent depairing factor $\Gamma/\Delta_0$ at temperature $T=0.01\,\mathrm{K}$. Blue line, right axis: plot of Ti scaled superconducting order parameter $\Delta/\Delta_0$ against scaled supercurrent depairing factor $\Gamma/\Delta_0$ at temperature $T=0.01\,\mathrm{K}$. Middle figure: Plot of critical current in reduced units $[I/I_0]^{2/3}$ against critical temperature in reduced units $T_c/T_{c,0}$ for Ti strips with thickness $t=100\,\mathrm{nm}$. (a) Black, dashed line - theoretical calculations; (b) red, cross markers - experimentally measured results for strip with width $w=1\,\mathrm{\mu m}$; (c) blue, square markers - experimentally measured results for strip with width $w=3\,\mathrm{\mu m}$. Right figure: Plot of critical current in reduced units $[I/I_0]^{2/3}$ against critical temperature in reduced units $T_c/T_{c,0}$ for Al-Ti bilayers with Al thickness $t_{Al}=25\,\mathrm{nm}$ and Ti thickness $t_{Ti}=100\,\mathrm{nm}$. (a) Black, dashed line - theoretical calculations; (b) red, cross markers - experimentally measured results for strip with width $w=3\,\mathrm{\mu m}$; (c) blue, square markers - experimentally measured results for strip with width $w=4\,\mathrm{\mu m}$; (d) green, diamond markers - experimentally measured results for strip with width $w=5\,\mathrm{\mu m}$.}
\end{figure}

The left figure of Fig.~\ref{fig:Combined_Fig_2} shows a plot of scaled current density $j/j_0$ and scaled superconducting order parameter $\Delta/\Delta_0$ against scaled depairing factor $\Gamma/\Delta_0$ for temperature $T=0.01\,\mathrm{K}$. There exist a maximum current density $j_c$. The pair of values $(j_{c},T)$ marks out a curve on the phase diagram within which the material is in the superconducting state, and beyond which the material is in the normal state. At $T\approx 0\,\mathrm{K}$, $j_c\approx0.746 j_{0}$, where $j_{0}=\sqrt{{N_S\sigma_N\Delta_0^3}/{\hbar}}$. It's worth noting that when $j\neq0$, transition to normal state happens when $\Delta$ is non-zero. Computationally, this means that the small $\Delta, \theta$ approximation technique, commonly used to compute the $T_c$ of $j=0$ transitions \citenumns{Martinis_2000,Songyuan_2018_Tc}, cannot be applied for these $j\neq0$ transitions.

The middle figure of Fig.~\ref{fig:Combined_Fig_2} shows a plot of critical current in reduced units $[I/I_0]^{2/3}$ against critical temperature in reduced units $T_c/T_{c,0}$ for a Ti strip with thickness $t=100\,\mathrm{nm}$. The right figure of Fig.~\ref{fig:Combined_Fig_2} shows a similar plot for Al-Ti bilayers with Al thickness $t_{Al}=25\,\mathrm{nm}$ and Ti thickness $t_{Ti}=100\,\mathrm{nm}$. The y-axis is chosen to reflect the Ginzburg-Landau result in the small supercurrent limit \citenumns{Romijn_1982} which states that ${I}/{I_0}\propto\left(1-{T_c}/{T_{c,0}}\right)^{3/2}$.
For both plots, the dotted line shows the values obtained from theoretical calculations using the Usadel equations; the scattered markers show the experimentally measured values for different widths of superconducting lines. The physical parameters used to generate the theoretical lines are the same as those used in \citenumns{Songyuan_2018_Tc}. To convert from $j$ to $I$, we have used $I=jtw$, where the thickness $t$ is deduced from calibrated deposition time, and the width $w$ is part of the design of the deposition mask. As expected from the analysis of length scales above, within each plot, wider superconducting lines result in earlier deviation from the ideal theoretical calculations. Denote $I_{c,0}$ as the \textit{actual} critical current of a device at close to $0\,\mathrm{K}$ (not to be confused with $I_0$ which is the \textit{theoretical} critical current). For most devices, the experimental data demonstrate good agreement with the theoretical prediction at $I<I_{c,0}/2$. For the widest bilayer device, good agreement is still obtain at $I<I_{c,0}/3$. This range encapsulates the common operating current values for typical TWPAs and KIDs systems: current much smaller than $I_{c,0}$ is usually chosen to avoid the onset of high current dissipation, or to avoid resonator bifurcation \citenumns{Eom_2012,Gao_2014,Swenson_2013,de_Visser_2010}. In this study, we have chosen a conservative thickness of $100\,\mathrm{nm}$. We expect an even bigger range of agreement for thinner devices such as the coplanar waveguides studied in \citenumns{Eom_2012}, which have thickness $t=35\,\mathrm{nm}$.

\section{Conclusions}
We have presented a numerical routine for analysing the inductance nonlinearity of thin-film superconductors with respect to supercurrent. Our analysis routine is based on the Usadel equations, Nam's equations for complex conductivity, transfer matrix calculation for complex surface impedances, and transmission line models. As appreciated in our discussion around the middle figure of Fig.~\ref{fig:Combined_Fig_1}, our analysis takes into account the full shape of superconducting densities of states and avoids an underestimation made by previous analyses on this subject. We have measured the superconducting transition temperature as a function of supercurrent for Ti single layers and Al-Ti bilayers. Our results show that the theory is in agreement with the experimental data in the current range that most thin-film superconductor devices are operated at, and therefore allows this analysis to be integrated in the design and optimization of future thin-film superconducting devices. Care needs to be taken when applying the numerical routine to AC applications, as both AC current distribution \citenumns{Withington_1995,Songyuan_transmission_lines_2018} and field quantization (in particular coherent excited states) \citenumns{deVisser_2014,Semenov_2016} effects are important at frequencies comparable to the superconductor pair-breaking frequency, and are likely to result deviation from the DC treatment in the Usadel equation formalism. Future studies should be conducted to investigate the extent of applicability as well as techniques to adapt the routine to AC applications.

\bibliographystyle{h-physrev}
\bibliography{library}
\end{document}